\documentclass[a4paper,11pt]{article}
\usepackage{graphics}
\usepackage{jheppub}
\usepackage{epsfig}
\usepackage{float}
\usepackage{slashed}
\expandafter\ifx\csname package@font\endcsname\relax\else
 \expandafter\expandafter
 \expandafter\usepackage
 \expandafter\expandafter
 \expandafter{\csname package@font\endcsname}%
\fi

\title{Muon anomalous magnetic moment and positron excess at AMS-02 in a gauged horizontal symmetric model}

\author[a,b]{Gaurav Tomar}
\author[a]{and Subhendra Mohanty}
\affiliation[a]{Physical Research Laboratory, Ahmedabad 380009, India.}
\affiliation[b]{Indian Institute of Technology, Gandhinagar 382424, India.}

\emailAdd{tomar@prl.res.in}
\emailAdd{mohanty@prl.res.in}

\abstract{ We studied an extension of the standard model with a fourth generation of fermions to explain the discrepancy in 
the muon $(g-2)$ and explain the positron excess seen in the AMS-02 experiment. We introduce a gauged $SU(2)_{HV}$ horizontal 
symmetry between the muon and the 4th generation lepton families. The 4th generation right-handed neutrino is identified as the 
dark matter with mass $\sim 700$ GeV. The dark matter annihilates only to $(\mu^+ \mu^-)$ and $(\nu^c_\mu ~\nu_\mu)$ states via 
$SU(2)_{HV}$ gauge boson. The $SU(2)_{HV}$ gauge boson with mass $\sim 1.4$ TeV gives an
adequate contribution to the $(g-2)$ of muon and fulfill the experimental constraint from BNL measurement. The higgs production
constraints from 4th generation fermions is evaded by extending the higgs sector.}

\begin{document}
\maketitle
\flushbottom
\section{Introduction}
There exist two interesting experimental signals namely the muon $(g-2)$, measured at BNL \cite{Bennett:2006fi, Bennett:2008dy} 
and the excess of positrons measured by AMS-02 \cite{5,Accardo:2014lma}, which may have a common beyond standard model (SM) 
explanation.\\ 
There is a discrepancy at $3.6\sigma$ level between the experimental measurement \cite{Bennett:2006fi, Bennett:2008dy} and 
the SM prediction \cite{Aoyama:2012wk,Gnendiger:2013pva,Davier:2010nc,Hagiwara:2011af,Benayoun:2012wc,
Blum:2013xva,Miller:2012opa} of muon anomalous magnetic moment,
\begin{equation}
 \Delta a_{\mu} \equiv a^{\rm Exp}_{\mu}-a^{\rm SM}_{\mu}= (28.7 \pm 8.0) \times 10^{-10},
 \label{er}
\end{equation}
where $a_{\mu}$ is the anomalous magnetic moment in the unit of $e/2m_{\mu}$.
In the standard model, contribution of $W$ boson to the muon anomalous magnetic magnetic moment goes as
$a^{W}_{\mu} \propto m^2_{\mu}/M^2_W$ and we have $a^{\rm SM}_{\mu}=19.48 \times 10^{-10}$ \cite{Beringer:1900zz}.\\ 
In minimal supersymmetric standard model (MSSM) \cite{Moroi:1995yh,Stockinger:2006zn}, we get contributions to muon $(g-2)$ from 
neutralino-smuon and chargino-sneutrino loops. In all MSSM diagrams there still exist a $m_{\mu}$ suppression in $(g-2)$, 
arising from the following cases:
(a) In case of bino in the loop, the mixing between the left and right handed smuons is $\propto m_{\mu}$
(b) In case of wino-higgsino or bino-higgsino in the loop, the higgsino coupling with smuon is $\propto y_{\mu}$, so
there is a $m_{\mu}$ suppression
(c) In the case of chargino-sneutrino in the loop, the higgsino-muon coupling is $\propto y_{\mu}$, which again gives rise 
to $m_{\mu}$ suppression. Therefor in MSSM $a^{\rm MSSM}_{\mu} \propto m^2_{\mu}/M^2_{\rm SUSY}$, 
where $M_{\rm SUSY}$ is proportional to the mass of the SUSY particle in the loop.\\ 
One can evade the muon mass suppression in $(g-2)$ with a horizontal gauge symmetry. In \cite{Baek:2001kca} a horizontal
$U(1)_{L_{\mu}-L_{\tau}}$ symmetry was used in which muon $(g-2)$ is proportional to $m_{\tau}$ and 
$a_{\mu}\propto m_{\mu}m_{\tau}/m^2_{Z^{\prime}}$, where $L_{\mu}-L_{\tau}$ gauge boson mass $m_{Z^{\prime}}\propto$ 
100 GeV gives the required $a_{\mu}$.
A model independent analysis of the beyond SM particles which can give a contribution to 
$a_{\mu}$ is studied in \cite{Queiroz:2014zfa}. The SM extension needed
to explain muon $(g-2)$ has also been related to dark matter \cite{Agrawal:2014ufa,Bai:2014osa} and the implication of this 
new physics in LHC searches has been studied \cite{Freitas:2014pua}. An explanation of $(g-2)$ from the 4th generation leptons 
has also been given in \cite{Carone:2013uh,BarShalom:2011bb}.\\
The second experimental signal, which we address in this paper is the excess of positron over cosmic-ray background, which has
been observed by AMS-02 experiment \cite{5} upto energy $\sim 425$ GeV \cite{Accardo:2014lma}. An analysis of
AMS-02 data suggests that a dark matter (DM) annihilation interpretation would imply that the annihilation final states are 
either $\mu$ or $\tau$ \cite{DeSimone:2013fia,Das:2013jca}. The dark matter annihilation into $e^{\pm}$ pairs would give a peak in 
positron signal, which is not seen in the positron spectrum. 
The branching ratio of $\tau$ decay to $e$ is only $17\%$ compared to $\mu$, which makes $\mu$ as the preferred source as
origin of high energy positrons. The AMS-02 experiment does not observe an excess, beyond the cosmic-ray background, in the 
antiproton flux \cite{Adriani:2010rc,AMS-02}, indicating a leptophilic dark matter \cite{Baek:2008nz,Dev:2013hka,4}.\\
In this paper, we introduce a 4th generation of fermions and a $SU(2)_{HV}$ vector gauge symmetry between the 4th generation 
leptons and the muon families. In our model, the muon $(g-2)$ has a contribution from the 
4th generation charged lepton $\mu^\prime$, and the $SU(2)_{HV}$ gauge boson $\theta^+$,
\begin{equation}
 \Delta a_{\mu} \propto \frac{ m_{\mu} m_{\mu^\prime}}{M^2_{\theta^+}}
\end{equation}
and from the neutral higgs scalars $(h,A)$,
\begin{equation}
 \Delta a_{\mu} \propto \frac{ m_{\mu}}{m_{\mu^\prime}}
\end{equation}
and from the charged higgs $H^\pm$ the contribution is,
\begin{equation}
 \Delta a_{\mu} \propto -\frac{ m_{\mu} m_{\nu_{\mu^\prime}}}{m^2_{H^\pm}}
\end{equation}
In all these cases, there is no quadratic suppression $\propto m^2_\mu$ because of the horizontal symmetry. By choosing parameters
of the model without any fine tunning, we can obtain the required number $\Delta a_\mu =2.87 \times 10^{-9}$ within $1\sigma$.
\\
In this model, the 4th generation right-handed neutrino $\nu_{\mu^\prime R}$, is identified as dark matter. The dark matter annihilates 
to the standard model particles through the $SU(2)_{HV}$ gauge boson $\theta_3$ and with the only final states being $(\mu^+ \mu^-)$
and $(\nu^c_\mu~\nu_\mu)$. The stability of DM is maintained by taking the 4th generation charged lepton to be heavier 
than DM. To explain the AMS-02 signal \cite{5,Accardo:2014lma}, one needs a cross-section (CS), 
$\sigma v_{\chi\chi\rightarrow\mu^+\mu^-} = 2.33 \times 10^{-25}\rm cm^3/sec$, which is larger than the CS, 
$\sigma v_{\chi\chi\rightarrow SM} \sim 3\times 10^{-26}\rm cm^3/sec$, required to get the correct thermal relic density
$\Omega h^2=0.1199\pm 0.0027$ \cite{6,7}. In our model, the enhancement of annihilation CS of DM in the galaxy is achieved by
the resonant enhancement mechanism \cite{2,3,4}, which we attain by taking $M_{\theta_3}\simeq 2m_\chi$.\\
This paper is organized as follows: In Sec.\ref{sec-II}, we describe the model. In Sec.\ref{sec-III} we discuss the 
dark matter phenomenology and in Sec.\ref{sec-IV}, we compute the $(g-2)$ contributions from this model and then give our 
conclusion in Sec.\ref{sec-V}.
\section{Model}\label{sec-II}
In addition to the three generations of quarks and leptons, we introduce the 4th generation of quarks $(c^\prime, s^\prime)$ and 
leptons $(\nu^\prime_\mu, \mu^\prime)$ (of both chiralities) in the standard model. We also add three right-handed neutrinos and extend
the gauge group of SM by horizontal symmetry denoted by $SU(2)_{HV}$, between the 4th generation lepton and muon families. Addition of 
three right-handed neutrinos ensures that the model is free from $SU(2)$ Witten anomaly \cite{Witten:1982fp}. We assume that the quarks
of all four generations and the leptons of $e$ and $\tau$ families are singlet of $SU(2)_{HV}$ to evade the constraints from flavour
changing processes. The $SU(2)_{HV}$ symmetry can be extended to $e$ and $\tau$ families by choosing suitable discrete symmetries, 
however in this paper we have taken $e$ and $\tau$ families to be singlet of $SU(2)_{HV}$ for simplicity and discuss the most 
economical model, which can explain muon $(g-2)$ and AMS-02 positron excess at the same time.\\
We denote the left-handed muon and 4th generation lepton families by
$\Psi_{Li\alpha}$ and their right-handed charged and neutral counterparts by $E_{R\alpha}$ and $N_{R\alpha}$ respectively
(here $i$ and $\alpha$ are the $SU(2)_L$ and $SU(2)_{HV}$ indices respectively and run through the values 1 and 2). 
The left-handed electron and tau doublets are denoted by $\psi_{eLi}$ and $\psi_{\tau Li}$ and their right-handed counterparts by $e_R$
and $\tau_R$ respectively. The gauge fields of $SU(2)_L\times U(1)_Y\times SU(2)_{HV}$ groups are denoted by 
$A^a_\mu,B_\mu$ and $\theta^a_\mu~(a=1,2,3)$ with gauge couplings $g,g^\prime$ and $g_H$ respectively.\\ 
The leptons transformations under the gauge group, 
$SU(3)_c \times SU(2)_L \times U(1)_Y \times SU(2)_{HV}\equiv G_{STD}\times SU(2)_{HV}$ are shown in Table.(\ref{qn}). 
From the assigned quantum numbers, it is clear that 
\begin{table}[ht!]
\begin{center}
\begin{tabular}{|c||c|c|}\hline
Particles & $G_{STD}\times SU(2)_{HV} ~\rm Quantum~numbers$\\\hline 
$\psi_{eLi} \equiv (\nu_e,e)$ & $(1,2,-1,1)$ \\\hline
$\Psi_{Li\alpha} \equiv (\psi_\mu,\psi_{\mu^\prime})$ & $(1,2,-1,2)$\\\hline
$\psi_{\tau Li} \equiv (\nu_\tau,\tau)$ & $(1,2,-1,1)$ \\\hline
$E_{R\alpha} \equiv (\mu_R,\mu^\prime_R)$ & $(1,1,-2,2)$ \\\hline
$N_{R\alpha} \equiv (\nu_{\mu R},\nu_{\mu^\prime R})$ & $(1,1,0,2)$ \\\hline
$e_R,\tau_R$ & $(1,1,-2,1)$ \\\hline
$\nu_{e R},\nu_{\tau R}$ & $(1,1,0,1)$\\\hline
$\phi_i$ & $(1,2,1,1)$ \\\hline
$\eta^\beta_{i\alpha}$ & $(1,2,1,3)$\\\hline
$\chi_\alpha$ & $(1,1,0,2)$ \\\hline
\end{tabular}
\end{center}
\caption{Representation of the various fields in the model under the gauge group $G_{STD}\times SU(2)_{HV}$.}
\label{qn}
\end{table}
the $SU(2)_{HV}$ gauge bosons connect only the leptons pairs, $\psi_{\mu_L}\leftrightarrow\psi_{\mu^\prime_L} {~\rm and~}
(\mu_R,\nu_{\mu R})\leftrightarrow(\mu^\prime_R,\nu^\prime_{\mu R})$. This assignment prevents the flavour changing process like 
$\mu\rightarrow e\gamma$ for which there are stringent bounds, and also ensures the contribution of heavy lepton
$\mu^\prime$ to the muon $(g-2)$ as shown in Fig.(\ref{fig5:mm}).
In our $G_{STD}\times SU(2)_{HV}$ model, the gauge couplings of the muon and 4th generation lepton families are,
\begin{align}\nonumber
 {\cal L}_{\psi} &=
   i\bar \Psi_{Li\alpha}\gamma^\mu
 \left(\partial_\mu-\frac{i}{2}g\tau \cdot A_\mu+ ig^\prime B_\mu-\frac{i}{2}g_H\tau \cdot \theta_\mu\right)_{ij;\alpha\beta}
 \Psi_{Lj\beta}\\\nonumber
 &+ i\bar E_{R\alpha}\gamma^\mu
 \left(\partial_\mu+ i2 g^\prime B_\mu-\frac{i}{2}g_H\tau \cdot \theta_\mu\right)_{\alpha\beta}E_{R\beta}
 + i\bar N_{R\alpha}\gamma^\mu
 \left(\partial_\mu-\frac{i}{2}g_H\tau \cdot \theta_\mu\right)_{\alpha\beta}N_{R\beta}\\
\end{align}
The ``neutral-current'' of $SU(2)_{HV}$ contributes to the annihilation process, 
$(\nu_{\mu^\prime}\nu_{\mu^\prime})\rightarrow\theta^*_3\rightarrow (\mu^+\mu^-),(\nu^c_\mu ~\nu_\mu)$, which is relevant for the 
AMS-02 and relic density calculations. The ``charge-changing'' vertex $\mu\mu^\prime \theta^+$, contributes to the $(g-2)$ of the muon.\\
To evade the bounds on the 4th generation from the higgs production at LHC, we extend the higgs sector (in addition to $\phi_i$) by a scalar 
$\eta^\beta_{i\alpha}$, which is a doublet under $SU(2)$ and triplet under $SU(2)_{HV}$. As a $SU(2)$ doublet $\eta^\beta_{i\alpha}$
evades 4th generation bounds from the overproduction of higgs in the same way as \cite{BarShalom:2011zj,He:2011ti}, in that the 
$125$ GeV mass eigenstate is predominantly $\eta$ which has no Yukawa couplings with the quarks. 
As $\eta^\beta_{i\alpha}$ is a triplet under $SU(2)_{HV}$, its Yukawa couplings with the muon and 4th generation lepton families 
split the masses of the muon and 4th generation leptons.
We also introduce a $SU(2)_{HV}$ doublet $\chi_\alpha$, which generates masses for $SU(2)_{HV}$ gauge bosons. The quantum 
numbers of the scalars are shown in Table.(\ref{qn}). The general potential
of this set of scalars $(\phi_i,\eta^\beta_{i\alpha},\chi_\alpha)$ is given in \cite{Ong:1980yk}. Following \cite{Ong:1980yk}, we
take the vacuum expectation values (vevs) of scalars as,
\begin{align}\label{vevs}\nonumber
 \langle \phi_i\rangle &= \langle \phi \rangle \delta_{i2},\\
 \langle \eta^\beta_{i\alpha}\rangle &= 
 \langle\eta\rangle \delta_{i2}(\delta_{\alpha1}\delta^{\beta1}-\delta_{\alpha2}\delta^{\beta2})\\\nonumber
 \lvert\langle\chi\rangle\rvert^2 &= \lvert\langle\chi_1\rangle\rvert^2 + \lvert\langle\chi_2\rangle\rvert^2 
\end{align}
where $\langle \phi_i\rangle$ breaks $SU(2)_L$, $\langle\chi_\alpha\rangle$ breaks $SU(2)_{HV}$ and $\langle\eta^\beta_{i\alpha}\rangle$
breaks both $SU(2)_L$ and $SU(2)_{HV}$ and generate the TeV scale masses for $SU(2)_{HV}$ gauge bosons. The mass eigenstates of the 
scalars will be a linear combination of $\phi_i,\eta^\beta_{i\alpha}$ and $\chi_\alpha$. We shall assume that the lowest mass
eigenstate $h_1$ with the mass $\sim 125$ GeV is primarily constituted by $\eta^\beta_{i\alpha}$. We shall also assume that the 
parameters of the higgs potential \cite{Ong:1980yk} are tuned such that mixing between $h_1$ and $\phi_i$ is small,
\begin{equation}
 \langle h_1|\phi_i\rangle \simeq 10^{-2},
\end{equation}
The Yukawa couplings of 4th generation quarks are only with $\phi_i$, therefore the $125$ GeV Higgs will have very small contribution 
from the 4th generation quarks loop.\\
The gauge couplings of the scalar fields $\phi_i, \eta^\beta_{i\alpha}$ and $\chi_\alpha$ are given by the Lagrangian,
\begin{align}\nonumber
{\cal L}_s&= \lvert (\partial_\mu - \frac{i}{2}g\tau \cdot A_\mu - ig^\prime B_\mu)\phi \rvert^2
            + \lvert (\partial_\mu - \frac{i}{2}g\tau \cdot A_\mu - ig^\prime B_\mu - ig_H T\cdot \theta_\mu)\eta \rvert^2\\
           & +  \lvert (\partial_{\mu} -\frac{i}{2}g_H\tau \cdotp \theta)\chi\rvert^2
\end{align}
where $\tau_a/2 ~(a=1,2,3)$ are $2\times2$ matrix representation for the generators of $SU(2)$ and $T_a ~(a=1,2,3)$ are $3\times3$
matrix representation for the generators of $SU(2)$. After expanding ${\cal L}_s$ around the vevs defined in Eq.(\ref{vevs}), the
masses of gauge bosons come,
\begin{equation*}
 M^2_W = \frac{g^2}{2}(2\langle\eta\rangle^2 + \langle\phi\rangle^2),~~M^2_Z = 
          \frac{g^2}{2} {\rm sec}^2\theta_W  (2\langle\eta\rangle^2 + \langle\phi\rangle^2),
 ~~M^2_A = 0,
 \end{equation*}
 \begin{equation}
  M^2_{\theta^+} = g_H^2 (4\langle\eta\rangle^2 + \frac{1}{2}\langle\chi\rangle^2),
  ~~M^2_{\theta_3} = \frac{1}{2} g^2_{H} \langle\chi\rangle^2
\label{masses}
\end{equation}
we tune the parameters in the potential such that the vevs of scalars are,
\begin{align}\nonumber
 2\langle\eta\rangle^2 + \langle\phi\rangle^2 &= (174~\rm GeV)^2\\
 \langle\chi\rangle &= 22.7~\rm TeV
 \label{nvevs}
 \end{align}
for the generation of large masses for 4th generation leptons $\mu^\prime,\nu_{\mu^\prime}$ and $SU(2)_{HV}$ gauge bosons 
$\theta^+, \theta_3$. The Yukawa couplings of the leptons are given by,
\begin{align}\nonumber
 {\cal L}_Y &= -h_1 \bar \psi_{eLi} \phi_ie_R - \tilde h_1\epsilon_{ij}\bar \psi_{eLi} \phi^j \nu_{eR}
 -h_2 \bar \Psi_{Li\alpha}\phi_i E_{R\alpha}-\tilde h_2\epsilon_{ij}\bar \Psi_{Li\alpha}\phi^j N_{R\alpha}
 -k_2 \bar \Psi_{Li\alpha}\eta^\beta_{i\alpha} E_{R\beta}\\
 &-\tilde k_2 \epsilon_{ij}\bar \Psi_{Li\alpha}\eta^{j\beta}_\alpha N_{R\beta} 
 -h_3 \bar \psi_{\tau Li} \phi_i \tau_R - \tilde h_3\epsilon_{ij}\bar \psi_{\tau Li} \phi^j \nu_{\tau R}+ \rm h.c 
\end{align}
after corresponding scalars take their vevs as defined in Eq.(\ref{vevs}), we obtain
\begin{align}\nonumber
  {\cal L}_Y &= -h_1 \bar \psi_{eL2}\langle\phi\rangle e_R - \tilde h_1 \bar \psi_{eL1} \langle\phi\rangle \nu_{eR}
  -\bar \Psi_{L2\alpha}[h_2 \langle\phi\rangle 
  + k_2\langle\eta\rangle(\delta_{\alpha1}-\delta_{\alpha2})]E_{R\alpha}\\\nonumber
  & -\bar \Psi_{L1\alpha}[\tilde h_2 \langle\phi\rangle 
  +\tilde k_2\langle\eta\rangle(\delta_{\alpha1}-\delta_{\alpha2})]N_{R\alpha}
  -h_3 \bar \psi_{\tau L2}\langle\phi\rangle \tau_R - \tilde h_3 \bar \psi_{\tau L1} \langle\phi\rangle \nu_{\tau R}\\\nonumber
  &-h_1 \bar \psi_{eLi} \phi^\prime_i e_R- \tilde h_1\epsilon_{ij}\bar \psi_{eLi} \phi^{\prime j} \nu_{eR}
  -\bar \Psi_{Li\alpha}[h_2 \phi^\prime_i\delta^\beta_\alpha  
  + k_2 \eta^{\prime\beta}_{i\alpha}]E_{R\beta}\\
  &-\bar \Psi_{Li\alpha}[\tilde h_2 \epsilon_{ij} \phi^{\prime j}\delta^\beta_\alpha  
  + \tilde k_2 \epsilon_{ij} \eta^{\prime j \beta}_{\alpha}]N_{R\beta}
  -h_3 \bar \psi_{\tau Li} \phi^\prime_i \tau_R- \tilde h_3\epsilon_{ij}\bar \psi_{\tau Li} \phi^{\prime j} \nu_{\tau R}+ \rm h.c
  \label{yukawas}
 \end{align}
where $\phi^\prime_i$ and $\eta^{\prime \beta}_{i\alpha}$ are the shifted fields. From Eq.(\ref{yukawas}), we see that
the muon and 4th generation leptons masses get split and are given by,
\begin{align}\nonumber
 m_e &= h_1 \langle\phi\rangle,
 ~~m_\tau = h_3 \langle\phi\rangle,
  ~~m_{\nu_e} = \tilde h_1 \langle\phi\rangle, 
 ~~m_{\nu_\tau} = \tilde h_3 \langle\phi\rangle\\
 ~~m_\mu &= h_2 \langle\phi\rangle + k_2 \langle\eta\rangle,
 ~~ m_{\nu_\mu} = \tilde h_2 \langle\phi\rangle + \tilde k_2 \langle\eta\rangle,\\
 m_{\mu^\prime} &= h_2 \langle\phi\rangle - k_2 \langle\eta\rangle,
 ~~m_{\nu_{\mu^\prime}} = \tilde h_2 \langle\phi\rangle - \tilde k_2 \langle\eta\rangle,\nonumber
 \label{mass}
 \end{align}
Thus by choosing the suitable values of Yukawas, the required leptons masses can be generated.
\begin{table}[ht!]
\begin{center}
\begin{tabular}{|c||c|c|}\hline
Parameters & Numerical values\\\hline 
$g_H$ & $0.087$ \\\hline
$y_h$ & $0.037$ \\\hline
$y_A$ & $0.020$ \\\hline
$y_{H^\pm}$ & $0.1$ \\\hline
$m_{\chi}$ & $700$ GeV\\\hline
$m_{\mu^\prime}$ & $740$ GeV \\\hline
$M_{\theta_3}$ & $1400$ GeV \\\hline
$M_{\theta^+}$ & $1400$ GeV \\\hline
$m_{H^\pm}$ & $1700$ GeV \\\hline
$m_h$ & $125$ GeV\\\hline
$m_A$ & $150~\mbox{GeV}$ \\\hline
$\delta$ & $10^{-3}$\\\hline
$\gamma$ & $10^{-4}$ \\\hline
\end{tabular}
\end{center}
\caption{Bench mark set of values used in the model.}
\label{pnv}
\end{table}
\section{Dark Matter Phenomenology}\label{sec-III}
\begin{figure}[t!]
\vspace*{10 mm}
\begin{center}
\includegraphics[scale=0.77,angle=0]{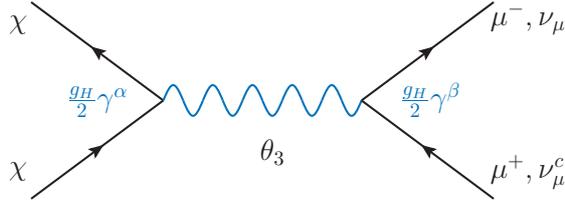}
\vspace*{3mm}
\caption{Feynman diagram of dark matter annihilation with corresponding vertex factor.}
\label{fig2}
\end{center}
\end{figure}

In our model, we identify the 4th generation right-handed neutral lepton $(\nu^\prime_{\mu_R}\equiv\chi)$ as the dark matter, 
which is used to fit AMS-02 data \cite{5,Accardo:2014lma}. The only possible channels for DM annihilation are into 
$(\mu^+\mu^-)$ and $(\nu^c_\mu~\nu_\mu)$ pairs (Fig.\ref{fig2}). In this scenario for getting the correct relic density, 
we use the Breit-Wigner resonant enhancement \cite{2,3,4} and take $M_{\theta_3} \simeq 2m_{\chi}$. 
The annihilation CS can be tuned to be $\sim 10^{-26}\rm cm^3 s^{-1}$ with the resonant enhancement, which gives the observed relic 
density. In principle the dark matter can decay into the light leptons via $SU(2)_{HV}$ gauge boson $\theta^+$
and scalar $\eta^\beta_{i\alpha}$, but by taking the mass of 4th generation charged leptons $\mu^\prime$ larger than $\chi$, the 
stability of dark matter can be ensured.
%
\subsection{Relic density}
The dark matter annihilation channels into standard model particles are, 
$\chi \chi \rightarrow \theta^\ast_3 \rightarrow \mu^+\mu^-,\nu^c_\mu\nu_\mu$. 
The annihilation rate of dark matter $\sigma v$, for a single channel, in the limit of massless leptons, is given by
\begin{equation}
 \sigma v = \frac{1}{16\pi} \frac{g^4_H m^2_\chi}{(s-M^2_{\theta_3})^2 + \Gamma^2_{\theta_3} M^2_{\theta_3}}
 \label{tcs}
\end{equation}
where $g_H$ is the horizontal gauge boson coupling, $m_\chi$ the dark matter mass, 
$M_{\theta_3}$ and $\Gamma_{\theta_3}$  are the mass and 
the decay width of $SU(2)_{HV}$ gauge boson respectively. Since both of the final states $(\nu_\mu, \mu)$ contribute in the relic 
density, the cross-section of Eq.(\ref{tcs}) is multiplied by a factor of 2 for relic density computation. The contributions to the 
decay width of $\theta_3$ comes from the decay modes, $\theta_3\rightarrow \mu^+\mu^-,\nu^c_\mu\nu_\mu$. The total decay width is given by,
\begin{equation}
 \Gamma_{\theta_3} = \frac{2 g^2_H}{48\pi} M_{\theta_3}
 \label{dw}
\end{equation}
In the non-relativistic limit, $s = 4m^2_\chi(1+v^2/4)$, then by taking into account the factor of 2, Eq.(\ref{tcs}) simplifies as,
\begin{equation}
 \sigma v = \frac{2}{256\pi m^2_\chi} \frac{g^4_H}{(\delta + v^2/4)^2 + \gamma^2}
 \label{mtcs}
\end{equation}
where $\delta$ and $\gamma$ are defined as $M^2_{\theta_3}\equiv 4m^2_\chi(1-\delta)$, and 
$\gamma^2 \equiv \Gamma^2_{\theta_3}(1-\delta)/4m^2_\chi$. If $\delta$ and $\gamma$ are larger than $v^2\simeq(T/M_\chi)^2$, the usual 
freeze-out takes place, on the other hand if $\delta$ and $\gamma$ are chosen smaller than $v^2$ then there is a resonant enhancement
of the annihilation CS and a late time freeze-out. We choose $\delta\sim 10^{-3}$ and $\gamma\sim10^{-4}$, so that we have a resonant 
annihilation of dark matter. The thermal average of annihilation rate is given as \cite{2,3,4},
\begin{equation}
 \langle \sigma v\rangle(x) = \frac{1}{n^2_{EQ}}\frac{m_\chi}{{64}\pi^4 x} \int^{\infty}_{4m^2_\chi} 
  \hat \sigma(s)\sqrt{s} K_1~ \left(\frac{x\sqrt{s}}{m_\chi}\right) ds,
  \label{tac}
\end{equation}
where,
\begin{equation}
 n^2_{EQ} = \frac{g_i}{2\pi^2}\frac{m^3_\chi}{x}K_2(x),
 \label{nEQ}
 \end{equation}
 \begin{equation}
  \hat \sigma(s) = 2g^2_i m_\chi \sqrt{s-4 m^2_\chi}~ \sigma v,
  \label{gi}
\end{equation}
and where $x\equiv m_\chi/T$; $K_1(x)$, $K_2(x)$ represent the modified Bessel functions of second type 
and $g_i$ is the internal degree of freedom of DM particle. Using Eq.(\ref{mtcs}), Eq.(\ref{nEQ}) and 
Eq.(\ref{gi}) in Eq.(\ref{tac}), it can be written as, 
\begin{equation}
 \langle \sigma v\rangle(x) =\frac{g^4_H}{512 m^2_\chi}\frac{x^{3/2}}{\pi^{3/2}}
 \int^{\infty}_{0}\frac{\sqrt{z}~{\rm Exp}[-xz/4]}{(\delta + z/4)^2 + \gamma^2} dz
 \label{tamc}
\end{equation}
where $z \equiv v^2$. We solve the Boltzmann equation for $Y_\chi=n_\chi/s$,
\begin{equation}
 \frac{dY_\chi}{dx} = -\frac{\lambda(x)}{x^2}(Y^2_\chi(x)-Y^2_{\chi eq}(x))
\end{equation}
where
\begin{equation}
 \lambda(x) \equiv \left(\frac{\pi}{45}\right)^{1/2} m_\chi M_{Pl}\left(\frac{g_{\ast s}}{\sqrt{g_\ast}}\right)
 \langle\sigma v\rangle(x)
\end{equation}
and where $g_\ast$ and $g_{\ast s}$ are the effective degrees of freedom of the energy density and entropy density respectively,
with $\langle\sigma v\rangle$ given in Eq.(\ref{tamc}). We can write the $Y_\chi(x_0)$ at the present epoch as,
\begin{equation}
 \frac{1}{Y_\chi(x_0)} = \frac{1}{Y_\chi(x_f)} + \int^{x_s}_{x_f} dx \frac{\lambda(x)}{x^2}
\end{equation}
where the freeze-out $x_f$ is obtained by solving $n_\chi(x_f)\langle\sigma v\rangle = H(x_f)$. We find that $x_f\sim 30$ and 
the relic density of $\chi$ is given by,
\begin{equation}
 \Omega = \frac{m_\chi s_0 Y_\chi(x_0)}{\rho_c}
\end{equation}
where $s_0=2890~\rm cm^{-3}$ is the present entropy density and $\rho_c = h^2 1.9\times 10^{-29}~\rm gm/cm^3$ is the critical density. We
find that by taking $g_H = 0.087,~\delta \sim 10^{-3}$ and $\gamma \sim 10^{-4}$ in Eq.(\ref{tamc}), we obtain the correct relic
density $\Omega h^2=0.1199\pm 0.0027$, consistent with Planck \cite{6} and WMAP \cite{7} data. From $g_H$ and $\gamma$ we can fix 
$M_{\theta_3}\simeq1400$ GeV and $m_\chi \simeq \frac{1}{2} M_{\theta_3}\simeq 700$ GeV.
There is a large hierarchy between the fourth generation charged fermion mass and the other charged leptons masses.
We do not have any theory for the Yukawa couplings and we take the $m_{\mu^\prime}$ mass which fits best the AMS-02 positron
spectrum and muon $(g-2)$. A bench mark set of values used in this paper for the masses and couplings is given in 
Table.(\ref{pnv}).
\subsection{Comparison with AMS-02 and PAMELA data}
The dark matter in the galaxy annihilates into $\mu^+\mu^-$ and the positron excess seen at AMS-02 \cite{5,Accardo:2014lma} appears from the decay 
of muon. We use publicly available code PPPC4DMID \cite{8,9} to compute the positron spectrum $\frac{dN_{e^+}}{dE}$ from the decay of 
$\mu$ pairs for 700 GeV dark matter. We then use the GALPROP code \cite{10,11} for propagation, in which we take the annihilation rate 
$\sigma v_{\mu^+\mu^-}$, and the positron spectrum $\frac{dN_{e^+}}{dE}$ as an input to the differential injection rate,
\begin{equation}
 Q_{e^+}(E, \vec r) = \frac{\rho^2}{2m^2_{\chi}}\langle \sigma v\rangle_{\mu^+\mu^-}\frac{dN_{e^+}}{dE}
 \label{st}
\end{equation}
where $\rho$ denotes the density of dark matter in the Milky Way halo, which we take to be the NFW profile \cite{12},
\begin{equation}
\rho_{\rm NFW}=\rho_0 \frac{r_s}{r} \left(1+\frac{r}{r_s}\right)^{-2},~\rho_0=0.4 {~\rm
GeV/cm^3},~r_s=20{~\rm kpc},
\end{equation}
\begin{figure}[t!]
\vspace*{10 mm}
\begin{center}
\includegraphics[scale=1.15,angle=0]{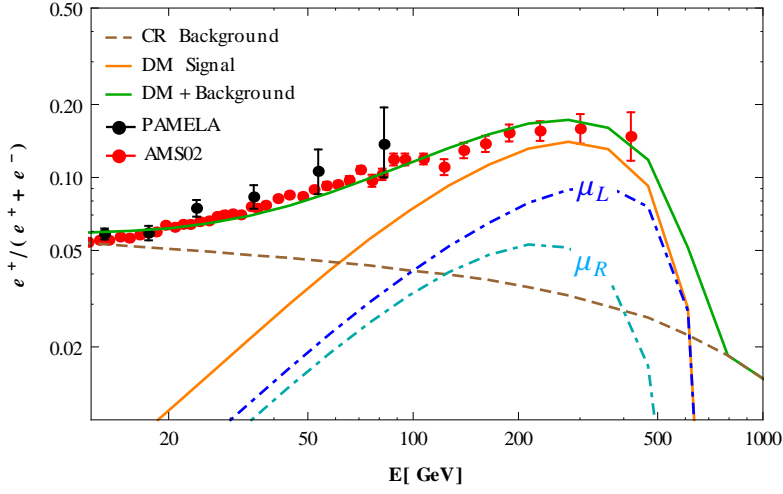}
\vspace*{3mm}
\caption{The positron flux spectrum compared with data from AMS-02 \cite{5,Accardo:2014lma} and PAMELA \cite{Adriani:2008zr}. 
The contributions of different channels $(\mu_L$, $\mu_R)$ are shown for comparison.}
\label{fig4}
\end{center}
\end{figure} 
In GALPROP code \cite{10,11}, we take the diffusion coefficient $D_0 = 3.6\times 10^{28}\rm cm^2 s^{-1}$ and Alfven speed 
$v_A =15~\rm Km s^{-1}$. We choose, $z_h=4~\rm kpc$ and $r_{max}=20 ~\rm kpc$, which are the half-width and maximum size for 
2D galactic model respectively. We choose the nucleus spectral index breaks at $9~\rm GeV$ and spectral index above this is 
$2.36$ and below is $1.82$. The normalization flux of electron
at $100~\rm GeV$ is $1.25\times 10^{-8}\rm cm^{-2} s^{-1} sr^{-1} GeV^{-1}$ and for the case of electron, we take
breaking point at $4~\rm GeV$ and its injection spectral index above $4~\rm GeV$ is $\gamma^{el}_1= 2.44$ and below 
$\gamma^{el}_0 = 1.6$. After solving the propagation equation, GALPROP \cite{10,11} gives 
the desired positron flux.\\
\begin{figure}[t!]
\vspace*{10 mm}
\begin{center}
\includegraphics[scale=1.15,angle=0]{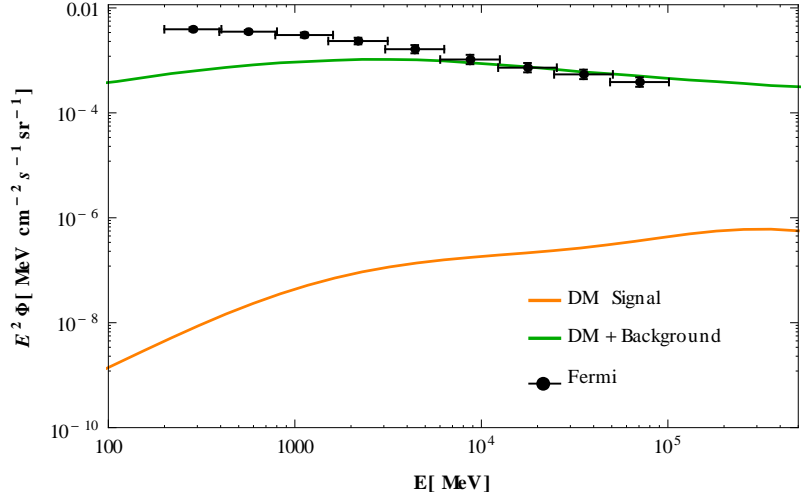}
\vspace*{3mm}
\caption{The $\gamma$-ray spectrum compared with data from Fermi Lat \cite{13}.}
\label{fig5}
\end{center}
\end{figure} 
To fit the AMS-02 data, the input annihilation CS required in GALPROP is, 
$\sigma v_{\chi\chi\rightarrow \mu^+\mu^-}=2.33\times 10^{-25}\rm cm^3 s^{-1}$.
The annihilation CS for $\mu$ final state from Eq.(\ref{tcs}) is, $\sigma v \approx 2.8 \times 10^{-25}\rm cm^3 s^{-1}$, which
signifies that there is no extra ``astrophysical'' boost factor needed to satisfy AMS-02 data. The annihilation rate required for relic
density was $\langle\sigma v\rangle \sim 3\times 10^{-26}\rm cm^3/sec$ and the factor $\sim 10$ increase in $\sigma v$ at the present
epoch is due to resonant enhancement by taking $m_\chi\simeq \frac{1}{2}M_{\theta_3}$.
In Fig.(\ref{fig4}), we plot the output of GALPROP code and compare it with the observed AMS-02 \cite{5,Accardo:2014lma} and PAMELA 
\cite{Adriani:2008zr} data. We see that our positron spectrum fits the AMS-02 data \cite{5,Accardo:2014lma} very well.
We also check the photon production from the 
decay of $\mu$ final state by generating the $\gamma$-ray spectrum called $\frac{dN_{\gamma}}{dE}$ from 
publicly available code PPPC4DMID \cite{8,9} and propagating it through the GALPROP code \cite{10,11}. We then compare the 
output with the observed Fermi-LAT data \cite{13}, as shown in Fig.(\ref{fig5}), and find that  
the  $\gamma$-ray does not exceed the observed limits. There is no annihilation to hadrons, so no excess of antiprotons are 
predicted, consistent with the PAMELA \cite{Adriani:2010rc} and AMS-02 \cite{AMS-02} data.
\section{Muon Magnetic Moment}\label{sec-IV}
\begin{figure}[t!]
\vspace*{10 mm}
\begin{center}
\includegraphics[scale=0.57,angle=0]{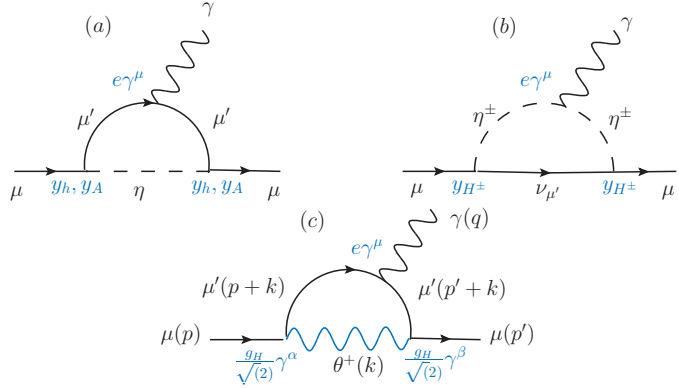}
\vspace*{3mm}
\caption{Feynman diagrams of scalar $\eta^\beta_{i\alpha}$ and $SU(2)_{HV}$ gauge boson $\theta^+$, which give contributions 
         to muon $(g-2)$.}
\label{fig5:mm}
\end{center}
\end{figure}

The $SU(2)_{HV}$ horizontal symmetry, which connects muon and 4th generation families, gives extra contributions to muon $(g-2)$. 
The diagrams that contribute to muon $(g-2)$ with $SU(2)_{HV}$ charged gauge boson 
$\theta^+$ and scalar $\eta^\beta_{i\alpha}$ are shown in Fig.(\ref{fig5:mm}).\\ 
We first calculate the contribution from $SU(2)_{HV}$ gauge boson $\theta^+$, which is shown in Fig.\ref{fig5:mm}(c). 
For this diagram the vertex factor of the amplitude $\mu(p^{\prime})\Gamma_{\mu}\mu(p)\epsilon^{\mu}$ is,
\begin{equation}
 \Gamma_{\mu} = \frac{e g^2_H}{2}\int \frac{d^4k}{(2\pi)^4} \gamma^{\beta}
                   \frac{(\slashed p^{\prime} + \slashed k + m_{\mu^\prime})}{(p^{\prime}+k)^2-m^2_{\mu^\prime}}\gamma_{\mu}
 \frac{(\slashed p + \slashed k + m_{\mu^\prime})}{(p+k)^2-m^2_{\mu^\prime}} \gamma^{\alpha} \frac{g_{\alpha\beta}}{k^2-M^2_{\theta^+}}
 \label{mc}
\end{equation}
we perform the integration and use the Gorden identity to replace, 
\begin{equation}
 (p_{\mu}+p^{\prime}_{\mu}) = 2m_{\mu}\gamma_{\mu} + i\sigma^{\mu\nu}q_{\nu}
\end{equation}
and identify the coefficient of the $i\sigma^{\mu\nu}q_{\nu}$ as the magnetic form factor. The contribution to $\Delta a_\mu$ is,
\begin{equation}
 [\Delta a_\mu]_{\theta^+} = \frac{m^2_\mu}{16\pi^2}\int^1_0 dx \frac{g^2_H\left(\frac{2m^\prime_\mu}{m_\mu}(x-x^2)-(x-x^3)\right)}
 {(1-x)m^2_{\mu^\prime}-x(1-x)m^2_\mu + x M^2_{\theta^\pm}}
\end{equation}
In the limit of $M^2_{\theta^+}>>m^2_{\mu^\prime}$, we get the anomalous magnetic moment,
\begin{equation}
 [\Delta a_{\mu}]_{\theta^+} =  \frac{g^2_H}{8 \pi^2}\left(\frac{m_{\mu}m_{\mu^\prime}-2/3 m^2_{\mu}}{M^2_{\theta^+}}\right)
 \label{nmm}
\end{equation}
we note that in Eq.(\ref{nmm}), the first term is dominant which shows $m_{\mu} m_{\mu^\prime}$ enhancement in the muon $(g-2)$. 
\\
In our model, the contribution from the neutral higgs $\eta$ (CP-even $h$ and CP-odd $A$) is shown in 
Fig.\ref{fig5:mm}(a). The $(g-2)$ contribution of this diagram is \cite{Leveille:1977rc},
\begin{align}\nonumber
 [\Delta a_\mu]_{h,A} &= \frac{m^2_\mu}{8\pi^2} \int^1_0 dx 
 \frac{y^2_h (x^2-x^3 + \frac{m_{\mu^\prime}}{m_{\mu}}x^2)}
 {m^2_\mu x^2 + (m^2_{\mu^\prime}-m^2_\mu)x + m^2_h(1-x)}\\
 & + \frac{m^2_\mu}{8\pi^2} \int^1_0 dx\frac{y^2_A (x^2-x^3 - \frac{m_{\mu^\prime}}{m_{\mu}}x^2)}
 {m^2_\mu x^2 + (m^2_{\mu^\prime}-m^2_\mu)x + m^2_A(1-x)}
 \label{spc}
\end{align}
where $y_h$, $y_A$ represent the Yukawa couplings of neutral CP-even and odd higgs respectively and their masses are denoted by $m_h$ and $m_A$
respectively. We shall calculate the contributions from the lightest scalars only, which give the larger contributions in 
compare to heavy scalars.
In the limits $m^2_{\mu^{\prime}}\gg m^2_h$, $m^2_{\mu^{\prime}}\gg m^2_A$, doing the integration in 
Eq.(\ref{spc}) we get the anomalous magnetic moment,
\begin{equation}
 [\Delta a_\mu]_{h,A} = \frac{1}{8\pi^2}\left(\frac{3 m_\mu m_{\mu^\prime} (y^2_h-y^2_A) + m^2_\mu(y^2_h+y^2_A)}{6m^2_{\mu^\prime}}\right)
\end{equation}
\\
In a similar way, the contribution from the mass eigenstate $H^\pm$ of charged higgs $\eta^\pm$, shown in 
Fig.\ref{fig5:mm}(b), is given by 
\cite{Leveille:1977rc},
\begin{equation}
 [\Delta a_\mu]_{H^{\pm}} = \frac{ m^2_\mu}{8\pi^2} \int^1_0 dx 
 \frac{y^2_{H^{\pm}} \left(x^3-x^2 + \frac{m_{\nu_{\mu^\prime}}}{m_{\mu}}(x^2-x)\right)}
 {m^2_\mu x^2 + (m^2_{H^{\pm}}-m^2_\mu)x + m^2_{\nu_{\mu^\prime}}(1-x)}
 \label{chc}
\end{equation}
where $y_{H^\pm}$ and $m_{H^\pm}$ are the Yukawa coupling and mass of the charged higgs respectively.
We perform the integration (Eq.\ref{chc}) in the limit $m^2_{H^\pm} \gg m^2_{\nu_{\mu^\prime}}$, and get the
anomalous magnetic moment,
\begin{equation}
 [\Delta a_\mu]_{H^{\pm}} = -\frac{y^2_{H^\pm}}{8\pi^2}\left(\frac{ 3 m_\mu m_{\nu_{\mu^\prime}}+m^2_\mu}{6m^2_{H^\pm}}\right)
\end{equation}
So the complete contribution to muon $(g-2)$ in our model is given as,
\begin{equation}
 \Delta a_\mu = [\Delta a_\mu]_{\theta^+} +  [\Delta a_\mu]_{h,A} + [\Delta a_\mu]_{H^{\pm}}
\end{equation}
As discussed before, in our model the lightest CP-even scalar $h_1$ is mainly composed of $\eta$, so we
can write,
\begin{equation}
y_h \sim k_2 ~cos \alpha_1
\end{equation}
where $\alpha_1$ is the mixing angle between CP-even mass eigenstate $h_1$ and gauge eigenstate $\eta$, and $k_2$ is the 
Yukawa coupling defined in Eq.(\ref{yukawas}). 
In the similar way, we assume that  lightest pseudoscalar $A$ and
charged higgs $H^\pm$ are also mainly composed of $\eta$, so that we can write
\begin{equation}
 y_A \sim k_2 ~cos \alpha_2,~~y_{H^\pm} \sim \tilde k_2 ~cos \alpha_3
\end{equation}
where $\alpha_2$ is the mixing angle between CP-odd scalars and $\alpha_3$ is the mixing angle between the charged scalars.
$\tilde k_2$ denotes the Yukawa coupling defined in Eq.(\ref{yukawas}).
\\
In the $SU(2)_H$ gauge boson sector, we take 
$g_H = 0.087,~M_{\theta^+} \approx 1400~\mbox{GeV}~(M_{\theta_3} \approx M_{\theta^+})$, which are fixed from the requirement
of correct relic density and we take $m_{\mu^\prime}= 740~\mbox{GeV}$, coming from the stability requirement of dark
matter $(m_{\mu^\prime} > m_\chi)$. After doing numerical calculation, we get $[\Delta a]_{\theta^+} = 3.61 \times 10^{-9}$.\\
The contribution from $(h,A)$ scalars depend on the parameter $k^2_2~(cos^2 \alpha_1 -cos^2 \alpha_2)$, which we assume to be
$\simeq 10^{-3}$ and obtain $[\Delta a_\mu]_{h,A} = 0.82 \times 10^{-9}$. For the charged scalar contribution, we assume 
$\tilde k_2 ~cos \alpha_3 = 0.1$ and $m_{H^\pm} = 1700$ GeV and obtain $[\Delta a_\mu]_{H^\pm} = -1.53 \times 10^{-9}$.
Adding the contributions from $\theta^+,~(h,A)$ and $H^\pm$, we get
\\
\begin{equation}
 \Delta a_\mu = 2.9 \times 10^{-9}
\end{equation}
which is in agreement with the experimental result \cite{Bennett:2006fi,Bennett:2008dy} within $1 \sigma$.
To get the desired value of muon $(g-2)$, we have to consider a large hierarchy between the neutral higgs 
($m_h\sim 125$ GeV, $m_A \sim 150$ GeV) and the charged higgs $m_{H^\pm}\sim 1700$ GeV. These masses have to arise by appropriate
choices of the couplings in the higgs potential of $(\phi_i,~\eta^\beta_{i\alpha},~\chi_\alpha)$.
\section{Result and Discussion}\label{sec-V}
We studied a 4th generation extension of the standard model, where the 4th generation leptons interact with the muon family via
$SU(2)_{HV}$ gauge bosons. The 4th generation right-handed neutrino is identified as the dark matter. We proposed a common explanation
to the excess of positron seen at AMS-02 \cite{5,Accardo:2014lma} and the discrepancy between SM prediction 
\cite{Aoyama:2012wk,Gnendiger:2013pva,Davier:2010nc,Hagiwara:2011af,Benayoun:2012wc,Blum:2013xva,Miller:2012opa} and BNL 
measurement \cite{Bennett:2006fi, Bennett:2008dy} of muon $(g-2)$. The $SU(2)_{HV}$ gauge boson $\theta^+$ with 4th generation 
charged lepton $\mu^\prime$ and charged higgs $H^\pm$, give the required contribution to muon $(g-2)$ to satisfy the BNL measurement 
\cite{Bennett:2006fi, Bennett:2008dy} 
within $1\sigma$. The LHC constraints on 4th generation quarks is evaded by extending the higgs sector
as in \cite{BarShalom:2011zj,He:2011ti}. In our horizontal $SU(2)_{HV}$ gauge symmetry model, we also explain the preferential 
annihilation of dark matter to $\mu^+\mu^-$ channel over other leptons and predict that there is no antiproton excess, in 
agreement with PAMELA \cite{Adriani:2010rc} and AMS-02 \cite{AMS-02} data. Since the dark matter has gauge interactions only 
with the muon family at tree level, we can evade the bounds from direct detection experiments 
\cite{Aprile:2012nq,Akerib:2013tjd} based on scattering of dark matter with the first generation quarks.
 
\end{document}